# Exact solution of three-dimensional (3D) spinless fermions

Zhidong Zhang

Shenyang National Laboratory for Materials Science, Institute of Metal Research,

Chinese Academy of Sciences, 72 Wenhua Road, Shenyang, 110016, P.R. China

**Abstract**

The three-dimensional (3D) Ising model is mapped into a 3D spinless fermionic model by the Jordan-Wigner transformation. The exact solution of the 3D model for spinless fermions is derived analytically by performing a diagonalization process consisting of the Clifford algebraic approach, the Fourier transformation and the Bogoliubov transformation. The Clifford algebraic approach is the same as that developed for the 3D Ising model, using a time average within the Jordan-von Neumann-Wigner framework, a linearization procedure and a local gauge transformation. The formulas for eigenvalues, partition function, subsequent thermodynamic properties and critical behaviors are presented. The dimensionality and the topological phases are investigated. The present results for many spinless fermions in a 3D lattice are applicable for studying the mechanisms of magnetism, superfluid, superconductors and topological materials.

The corresponding author: Z.D. Zhang, Tel: 86-24-23971859, Fax: 86-24-23891320, e-mail address: zdzhang@imr.ac.cn

## 1. Introduction

Fermions and bosons are two fundamental (quasi-)particles in the nature, which possess different behaviors and follow different quantum statistics. Spinless fermions are particularly interesting since they are basic excitations in various physical systems, which control physical properties. Many interesting phenomena in condensed matter physics are closely related with spinless fermions, among which we mention the following issues, superconductors [1], fractional quantum Hall effect [2], topological insulators [3], Majorana fermions [4], quantum anomalous Hall insulators [5,6], and topological nodal fermion semimetals [7]. Therefore, it is extremely important to have a deep understanding on systems of interacting many fermions, and the exact solutions are quite helpful for this purpose. Up to date, none of such a solution has been reported for a three-dimensional (3D) model of spinless fermions.

The Ising model is one of the simplest models for describing many-body interactions between spins (or particles) [8]. It was uncovered by Schultz, Mattis and Lieb [9] that a two-dimensional (2D) Ising model serves as a soluble problem of many fermions. Their results for a (1+1)D spinless fermionic model (time is the second dimension) are consistent with the exact solution derived by Onsager [10] and Kaufman [11] for the 2D Ising model at zero external magnetic field. The Jordan-Wigner transformation [12], which transforms a many-body interacting spin system to a spinless fermionic system, has been applied to various physical systems [13-17]. Fradkin, Srednicki and Susskind showed that the 3D Ising model (and also the $Z_2$ lattice gauge theory is equivalent to a theory of locally interacting fermions, which turns out to describe a theory of surfaces [17].

The exact solution of the 3D Ising model is a well-known hard problem in physics. Two conjectures were proposed by the present author in [18] and then rigorously proven in collaboration with Suzuki and March [19-21] for solving the exact solution of the ferromagnetic 3D Ising model at zero magnetic field. The exact solution of the 3D Ising model provides an opportunity to derive the exact solutions of other related models, for instance, a 2D Ising model with a transverse field [22], a 3D $Z_2$ lattice gauge theory [23] and a (3+1)D $\phi^4$ scalar field model with ultraviolet cutoff [24]. In this work, we shall derive the exact solution of a 3D model for many spinless fermions.

The paper is organized as follows: In Section 2, the Hamiltonian for the 3D Ising model is transformed by the Jordan-Wigner transformation to that for a 3D model of spinless fermions. In Section 3, a diagonalization process is introduced, which includes the Clifford algebraic approach, the Fourier transformation and the Bogoliubov transformation. The Clifford algebraic approach consists of a time average within the Jordan-von Neumann-Wigner framework [25], a linearization procedure and a local gauge transformation. In Section 4, the eigenvalues, the partition function, subsequent thermodynamic properties and critical behaviors are investigated. In Section 5, the dimensionality and the topological phases are discussed. Section 6 is for conclusions.

**2. Hamiltonians**

The Hamiltonian of the 3D Ising model is represented as [8,18]:

$$H_0 = -\sum_{<i,j>}^{N} J_{ij} S_i S_j$$

(1)

Here, spins with $S = 1/2$ are arranged on a 3D lattice with lattice size $N = mnl$, where $m$, $n$, $l$ denote the lattice points along three crystallographic directions. For simplicity, only are the nearest neighboring interactions considered. For a ferromagnetic 3D Ising model, all the interactions $J_{ij}$ are positive, which are set to be $J_1$, $J_2$ and $J_3$ along the three directions

The Hamiltonian of the 3D Ising model is reduced:

$$H_1 = -K_1^* \sum_{i=1}^{N_1} \sigma_i^z - K_2 \sum_{i=1}^{N_1-1} \sigma_i^x \sigma_{i+1}^x - K_3 \sum_{i=1}^{N_1-1} \sigma_i^x \sigma_{i+n}^x$$

(2)

by the Jordan-Wigner transformation [12]. We have introduced $K_1 = \frac{J_1}{k_B T}$, $K_2 = \frac{J_2}{k_B T}$ and $K_3 = \frac{J_3}{k_B T}$, with the dual interaction $\tanh K_1^* = e^{-2K_1}$ [10,11,18]. In Eq. (2) for the Hamiltonian $H_1$, a factor of $(2\sinh 2K_1)^{\frac{mnl}{2}}$ is dropped temporarily, which should be picked up for calculating the partition function $Z$ [18,19]. The number of spins is taken to be $N_1 = nl$ for the sum of spin-spin interactions in a plane, because in the calculation process for the partition function $Z$, $m$ lattice sites along one dimension have been used already for the periodic condition to result in $\lambda_{i\prime}^m$ for eigenvalues [18,19].

We employ the Jordan-Wigner transformation [12]:

$$\sigma_i^z = (2c_i^\dagger c_i - 1)$$

(3)

$$\sigma_i^x = \prod_{j<i}(1 - 2c_j^\dagger c_j)(c_i + c_i^\dagger)$$

(4)

where $c_j^\dagger$ and $c_j$ are fermionic creation and annihilation operators, respectively. The Hamiltonian becomes the following form for spinless fermions [9,13]:

$$H_2 = -\sum_{i=1}^{N_1-1}\left[K_1^*(2c_i^\dagger c_i - 1) + K_2(c_i^\dagger - c_i)(c_{i+1} + c_{i+1}^\dagger) \right.$$
$$\left. + K_3(c_i^\dagger - c_i)\prod_{i<j<i+n}(1 - 2c_j^\dagger c_j)(c_{i+n} + c_{i+n}^\dagger)\right]$$

(5)

Some phase factors appear at the hopping terms in the Hamiltonian of many fermionic systems after the Jordan-Wigner transformation, due to the boundary conditions, high dimensions and/or the application of a magnetic field [26-36].

The boundary condition may cause some difference in the formulation of the problem. The boundary factors $U$ of the 2D Ising model split the space of the transfer matrices [11], which are neglected in the thermodynamic limit [19]. The eigenstates of transfer matrices for spinless fermions can be considered separately as states in which only even numbers of fermions are present, and other states in which only odd numbers of fermions are present [9]. We have omitted the phase factors corresponding to the boundary factors $U$ in the Hamiltonian $H_2$. The nonlinear terms in the Hamiltonian $H_2$ (Eq. (5)) are the same as the internal factors $W_j$ for every site $j$ in the 3D Ising model [19,20], which can be seen as Chern-Simons gauge fields acting on each site [16,37]. It has the same characters with the boundary factors $U$, splitting the Hilbert space of the system. But the effects of the internal factors $W_j$ cannot be neglected in the thermodynamic limit, because an internal factor appears on every lattice site $j$ [19].

**3. Diagonalization process**

In this section, we introduce the diagonalization process for the Hamiltonian $H_2$ (Eq. (5)) of the 3D spinless fermionic model, which can be carried out by the combination of the following transformations/approaches:

3.1 Clifford algebraic approach

The Clifford algebraic approach developed in [19] proved rigorously four theorems (*i.e.,* trace invariance, linearization, local transformation and commutation theorems) for the 3D Ising model. The Clifford algebraic approach consists of the following procedures:

3.1.1 Time average within Jordan-von Neumann-Wigner framework

The Jordan-von Neumann-Wigner theorem [25] provides the mathematical basis of quantum mechanics. It has been proven that models in the topological quantum statistical mechanics (such as the 3D Ising model) must be set up on the Jordan-von Neumann-Wigner framework, with application of Jordan algebra $A \circ B = \frac{1}{2}(AB + BA)$ for multiplication of operators, which overcomes the difficulty of noncommutative operators, and ensures the integrability of the many-body spin system [19]. It equalizes to perform a time average for a physical quantity, violating the ergodic hypothesis at finite temperatures [38], which provides a chance to integral the physical quantity along a closed path with a monodromy representation (or the parallel transport) [20]. This causes a gap existing between the initial and final states of the system with time evolution, represented by topological phases on the eigenvectors and the eigenvalues (and also the partition function) [18-21].

One of the difficulties for solving analytically the 3D Ising model is the anti-commutativity of fermionic operators, which leads to a factor $(-1)^F$ up to the $F$ times of commutations of two fermionic operators at different sites. This causes different states of the system, depending on even and odd times for commutations of the fermionic operators. In order to overcome such a difficulty, the first procedure we employ includes the time average, expanding the 3D system to be of (3+1)-dimensions with a transformation of wavefunction $\Psi_{3D} \to \Psi_{(3+1)D}$, while using Jordan algebra for multiplication of fermionic operators. Expanding the system and using Jordan algebra transform the Hamiltonian to a new form:

$$H_3 = -\sum_{i=1}^{N_2-1}\left[K_1^*(2c_i^{\dagger}\circ c_i - 1) + K_2(c_i^{\dagger} - c_i)\circ(c_{i+1} + c_{i+1}^{\dagger}) \right.$$
$$\left. + K_3(c_i^{\dagger} - c_i)\circ \prod_{i<j<i+n} \circ(1 - 2c_j^{\dagger}\circ c_j) \circ(c_{i+n} + c_{i+n}^{\dagger})\right]$$

(6)

Here the notation ° represents that the products between the fermionic operators follow the multiplication of Jordan algebra. For the Hamiltonian $H_3$, the number of spins is taken to be $N_2 = nlo$ for the sum of interactions between the nearest neighboring spins at lattice sites. Introducing the fourth direction with lattice points $o$ expands the system to be (3+1) dimensional with a much larger Hilbert space [18-21]. The Hilbert space of the 3D Ising system is expressed by the quaternion representation, in which the 2n-normalized eigenvectors (Eq. (54) of Kaufman [11]) are generalized to construct the 2nlo-normalized eigenvectors $\Psi_{(3+1)D}$. The eigenvectors $\Psi_{(3+1)D}$ are represented as Eq. (33) of ref. [18], but at the present stage without the weight factors $w_x$, $w_y$ and $w_z$. In the thermodynamic limit $n \to \infty$, $l \to \infty$, $o \to \infty$, we have the relation for the number of the spins: $N_1 = N_2^{2/3}$. The total energy for the system with $H_3$ is written as: $E_3 = N_2 \varepsilon_3$, where $\varepsilon_3$ represents three interaction terms in the sum of Eq. (6). Meanwhile, the total energy for the system with $H_2$ is described as: $E_2 = N_1 \varepsilon_2$, where $\varepsilon_2$ represents three interaction terms in the sum of Eq. (5). Since $\varepsilon_2 = \varepsilon_3$ and $N_1 = N_2^{2/3}$, in order to keep the invariance of the energy of the system, we must have the equivalent relation between the energies with $H_2$ and $H_3$: $E_2 = N_1 \varepsilon_2 = N_2^{2/3} \varepsilon_3 = E_3^{2/3}$. This conservation of energy, in accordance with the dimension analysis, will be held also for the energy-momentum relation. After we obtain the energy-momentum dispersion, we have to project the system from (3+1)-dimensions to three dimensions (see subsection 3.3).

3.1.2 Linearization procedure

For the linearization procedure [19], we first inspect the effect of the boundary factor $U$, which splits the space of the transfer matrices and the Hilbert space. The boundary factor is expressed as [30]:

$$e^{i\pi n_+} = exp\left( i\pi \sum_{j=1}^{n} c_j^\dagger c_j \right)$$

(7)

In a convenient way, two projecting cases can be simultaneously accounted for by introducing a matrix representation for fermionic operators. They are projected onto the subspaces with even and odd eigenvalues of $n_+$ by means of the projectors $P^\pm$ [19,30,39]:

$$P^{\pm} = \frac{1}{2}\left[1 \pm e^{i\pi n_+}\right]$$

(8)

with $P^+ + P^- = 1$. The Hamiltonian $H$ conserves the number of fermions, i.e., $[H, e^{i\pi n_+}] = 0$

$$H = \begin{pmatrix} P^+HP^+ & P^+HP^- \\ P^-HP^+ & P^-HP^- \end{pmatrix} \equiv \begin{pmatrix} H^{(+)} & 0 \\ 0 & H^{(-)} \end{pmatrix}$$

(9)

In order to solve the problem with nonlinear terms in the 3D spinless fermionic model, we generalize the linearization procedure of the boundary factor $U$ to linearize the internal factors $W_j$:

$$W_j = \prod_{i<j<i+n}(1 - 2c_j^{\dagger}\circ c_j) = \prod_{i<j<i+n} e^{i\pi c_j^{\dagger}\circ c_j}$$

(10)

The internal factors $W_j$ can be taken as the projection operator $P \equiv \frac{1}{2}[1 \pm W_j]$ [19,30,39], which projects the system from a larger Hilbert space with more number of states to another smaller Hilbert space with less number of states. The projection is to remove unphysical states and keep only physical states. Linearizing the nonlinear terms in the problem is realized by the projection operators $P$. It should be noticed that there are $2^{nl}$ terms of combinations of positive/negative signs, representing $2^{nl}$ subspaces [19]. According to the largest eigenvalue principle, only the largest eigenvalue contributes dominantly to the partition function $Z$ of the 3D Ising model in the thermodynamic limit [19]. We can choose the largest eigenvalue in one piece of subspaces among all the eigenvalues distributed in $2^{nl}$ subspaces, which possesses positive signs for all the terms for every site $i$. Subsequently, the

Hamiltonian corresponding to the largest eigenvalue is denoted as $H^{(++\cdots++)} = H_4$:

$$H_4 = -\sum_{i=1}^{N_2-1} \left[K_1^*(2c_i^{\dagger}\circ c_i - 1) + K_2(c_i^{\dagger} - c_i)\circ(c_{i+1} + c_{i+1}^{\dagger}) \right.$$
$$\left. + K_3(c_i^{\dagger} - c_i)\circ(c_{i+n} + c_{i+n}^{\dagger})\right]$$

(11)

The largest eigenvalue principle ensures that the partition functions for the systems with the Hamiltonians $H_3$ and $H_4$ are equivalent. Note that the terms with $K_1^*$, $K_2$ and $K_3$ in the Hamiltonian $H_4$ represent rotations along three crystallographic directions, respectively. The linearization procedure above is correct, because we can check that $\sigma_j^x \sigma_{j+n}^x$ and $\sigma_j^x \sigma_{j+1}^z \ldots \sigma_{j+n-1}^z \sigma_{j+n}^x$ commute and thus have the same eigenvalues. However, now the problem becomes how to fix the eigenvalues for $\sigma_j^x \sigma_{j+n}^x$ terms, which reply on the global effect of the whole physical system. Namely, all the terms in the Hamiltonian $H_4$ including the nontrivial topological structures with crossings contribute to the eigenvalues, which contain a global effect. A local gauge transformation is needed for this purpose.

3.1.3 Local gauge transformation

There still exists a problem to be dealt with, which is topological. It is clearly visible from the third term $(c_i^{\dagger} - c_i)\circ(c_{i+n} + c_{i+n}^{\dagger})$, which shows the nontrivial topological structures. According to the topological theory, the nontrivial topological structures contribute also to the partition function and the thermodynamic properties of the physical system. One has to perform a rotation to trivialize the nontrivial topological structures, transforming $(c_{i+n} + c_{i+n}^{\dagger})$ to $(c_{i+1} + c_{i+1}^{\dagger})$, while generalizing topological phases on eigenvectors of the system. A mapping exists between the crossings of the topological structures and the spin alignments [21]. The crossings of the topological structures effectively contribute to the free energy of the system the same as what the spin alignments do. The rotation matrix added in Eqs. (22) and (24) of [18] for trivializing the nontrivial topological structures represents a Lorentz transformation. The topological Lorentz transformation is acting as a gauge transformation, while generating topological phases on eigenvalues and eigenvectors. It equalizes to add an extra term in the Hamiltonian, because a virtual dimension exists in the spin lattice (see [21]), representing the rotation in the fourth dimension and standing for the topological contributions to physical properties [18-21]. Meanwhile, the topological phases emerge on the eigenvectors $\Psi_{(3+1)D}$ so that the eigenvectors become $\Psi'_{(3+1)D}$. The procedure is described in formulations as follows:

Perform a local gauge transformation $G$:

$$GH_4G^{-1} \equiv H_5 \to H_6$$

(12)

while,

$$G\Psi_{(3+1)D} \to \Psi'_{(3+1)D}$$

(13)

The local gauge transformation equalizes adding an interaction $K_4$ in the Hamiltonian $H_4$ as the representation of a rotation. The Hamiltonian $H_5$ is written as:

$$H_5 = -\sum_{i=1}^{N_2-1} \left[ K_1^*(2c_i^{\dagger}\circ c_i - 1) + K_2(c_i^{\dagger} - c_i)\circ(c_{i+1} + c_{i+1}^{\dagger}) \right.$$
$$\left. + K_3(c_i^{\dagger} - c_i)\circ(c_{i+n} + c_{i+n}^{\dagger}) + K_4(c_i^{\dagger} - c_i)\circ(c_{i+n'} + c_{i+n'}^{\dagger}) \right]$$

(14)

Here the interaction $K_4 = K_2 K_3 / K_1$ acts between the nearest neighboring spins along the fourth virtual dimension for the spin lattice [18]. It is determined by the star-triangle relation $K_1 K_1^* = K_1 K_2 + K_1 K_3 + K_2 K_3$ [18], which is the exact solution of the Yang-Baxter equation in the continuous limit [40]. The generalized Yang-Baxter equation (so-called tetrahedron equation) guarantees the integrability of the 3D Ising model (and also the 3D spinless fermionic model), via the topological transformation keeping the invariant of bracket polynomial (partition function) under the Reidemeister moves II and III [18-21]. The gauge transformation transforms the eigenvectors $\Psi_{(3+1)D}$ to the eigenvectors $\Psi'_{(3+1)D}$ for the 3D spinless fermionic model. These eigenvectors $\Psi'_{(3+1)D}$ are the same as those for the 3D Ising model, as represented in Eq. (33) of [18], in which the weight factors $w_x$, $w_y$, $w_z$ (i.e., topological phases) emerge. Simultaneously, the gauge transformation is acting on creation and annihilation fermion operators in the Hamiltonian $H_5$ [27,29]:

$$c_{i+1} \to e^{-i\phi_x} c_{i+1}$$
$$c_{i+1}^{\dagger} \to e^{i\phi_x} c_{i+1}^{\dagger}$$
$$c_{i+n} \to e^{-i\phi_y} c_{i+1}$$
$$c_{i+n}^{\dagger} \to e^{i\phi_y} c_{i+1}^{\dagger}$$

(15)

and

$$c_{i+n'} \to e^{-i\phi_z} c_{i+1}$$

$$c^\dagger_{i+n'} \to e^{i\phi_z} c^\dagger_{i+1}$$

(16)

The phase factors generated here are direct results of transformation between spins located at different sites $j+1$ and $j + n$ (or $j + n'$). For an operator at lattice site $i$, $c_i = e^{-i\varphi_i} S_i^-$, while for lattice site $i + n$, we have $c_{i+n} = e^{-i\varphi_{i+n}} S_{i+n}^-$ [27]. Then we have

$$\frac{c_{i+n}}{c_i} = e^{-i(\varphi_{i+n}-\varphi_i)} \frac{S_{i+n}^-}{S_i^-}$$

(17)

If we set $S_{i+n}^- \to S_i^-$, $c_{i+n} \to e^{-i(\varphi_{i+n}-\varphi_i)} c_i = e^{-i\Delta\varphi_{i+n}} c_i$. Thus, the Hamiltonian is simultaneously transformed to:

$$H_6 = -\sum_{i=1}^{N_2-1} \left[ K_1^*(2c_i^\dagger c_i - 1) + K_2(c_i^\dagger - c_i)(e^{-i\phi_x} c_{i+1} + e^{i\phi_x} c_{i+1}^\dagger) \right.$$
$$+ K_3(c_i^\dagger - c_i)(e^{-i\phi_y} c_{i+1} + e^{i\phi_y} c_{i+1}^\dagger)$$
$$\left. + K_4(c_i^\dagger - c_i)(e^{-i\phi_z} c_{i+1} + e^{i\phi_z} c_{i+1}^\dagger) \right]$$

(18)

Here we remove the notation ° for the multiplication of Jordan algebra, sine the difficulty of noncommutative operators has been overcome. In what follows, we shall use the normal multiplication instead of Jordan algebra, for multiplication of fermion operators.

3.2 Fourier transformation

The Fourier transformation is described as [9,13]:

$$c_i = \left(\frac{1}{M}\right)^{1/2} \sum_k e^{-ik \cdot r_i} c_k$$

$$c_i^\dagger = \left(\frac{1}{M}\right)^{1/2} \sum_k e^{i\mathbf{k}\cdot\mathbf{r}_i} c_k^\dagger$$

(19)

with $M = (2n+1)(2l+1)(2o+1)$. The Fourier transformation results in the Hamiltonian in the momentum space [9,13]:

$$H_7 = -2\sum_{k>0} A_k(c_k^\dagger c_k + c_{-k}^\dagger c_{-k}) + 2i\sum_{k>0} B_k\left[c_k^\dagger c_{-k}^\dagger + c_k c_{-k}\right]$$

(20)

with

$$A_k = K_1^* + K_2\cos(k_x a + \phi_x) + K_3\cos(k_y a + \phi_y) + K_4\cos(k_z a + \phi_z)$$

(21)

and

$$B_k = K_2\sin(k_x a + \phi_x) + K_3\sin(k_y a + \phi_y) + K_4\sin(k_z a + \phi_z)$$

(22)

Here the complete set of the wave vector ***k*** is described by three components ($k_x$, $k_y$, $k_z$) along three directions for the cyclic condition, which forms a quaternion form in the (3+1)D framework and $a$ is the lattice constant. We have the following geometric relation [13]:

$$tan2\theta_k = -\frac{B_k}{A_k}$$

(23)

3.3 Bogoliubov transformation

The Bogoliubov transformation to a new set of operators $\eta_k$ and $\eta_k^\dagger$ are represented as [9,13]:

$$c_k = u_k \eta_k - i v_k \eta^\dagger_{-k}$$

$$c_{-k} = u_k \eta_{-k} + i v_k \eta^\dagger_k$$

$$c^\dagger_k = u_k \eta^\dagger_k + i v_k \eta_{-k}$$

$$c^\dagger_{-k} = u_k \eta^\dagger_{-k} - i v_k \eta_k$$

(24)

Performing the Bogoliubov transformation realizes the Hamiltonian $H_8$ in the diagonalization form:

$$H_8 = \sum_k \Lambda'_k \eta^\dagger_k \eta_k + constant$$

(25)

The energy gap in $H_6$ is given by:

$$\Lambda'_k = 2\sqrt{A_k^2 + B_k^2}$$

(26)

However, according to $E_2 = E_3^{2/3}$ (see subsection 3.1.1) and the equivalences $H_1 = H_2$ and $H_3 = H_4 = H_5 = H_6 = H_7 = H_8$, keeping the number of the wave vector $\boldsymbol{k}$ the same, the relation $E_2(k) = E_8(k)^{2/3}$ is validated. Thus we have a new form of the Hamiltonian:

$$H_9 = \sum_k \Lambda_k \zeta^\dagger_k \zeta_k + constant$$

(27)

with

$$\Lambda_k = 2^{2/3} \cdot \sqrt[3]{A_k^2 + B_k^2}$$

(28)

for setting up the equivalence to the Hamiltonian $H_1$ and also to the original Hamiltonian $H_0$ in consideration of picking up a factor of $(2sinh2K_1)^{\frac{mnl}{2}}$ for the partition function $Z$ (see Section 2).

## 4. Eigenvalues, partition function, spontaneous magnetization, correlation and mass gap

In this section, we summarize the exact results for the eigenvalues, the partition function, the eigenvectors, the spontaneous magnetization, the true range of the correlation and the mass gap for the 3D spinless fermionic model.

The energy eigenvalues for the 3D many spinless fermions are written as [18]:

$$\cosh\gamma_{2t} = cosh2K_1^* \cdot cosh2(K_2 + K_3 + K_4) - sinh2K_1^* \cdot sinh2(K_2 + K_3 + K_4)$$
$$\times \left[\cos\left(\frac{2t_x\pi}{n} + \phi_x\right) + \cos\left(\frac{2t_y\pi}{l} + \phi_y\right) + \cos\left(\frac{2t_z\pi}{o} + \phi_z\right)\right]$$

(29)

The eigenvectors $\Psi'_{(3+1)D}$ are the same as Eq. (33) of ref. [18], but the weight factors $w_x$, $w_y$ and $w_z$ are replaced by the phase factors $e^{i\phi_x}$, $e^{i\phi_y}$ and $e^{i\phi_z}$. Meanwhile, the geometric relations are as those represented in Eqs. (30) and (31) of [18].

The partition function of the 3D spinless fermionic model is expressed as [18]:

$$N^{-1} \ln Z = \ln 2$$
$$+ \frac{1}{2(2\pi)^4} \int_{-\pi}^{\pi}\int_{-\pi}^{\pi}\int_{-\pi}^{\pi}\int_{-\pi}^{\pi} \ln[cosh2K_1 \cdot cosh2(K_2 + K_3 + K_4)$$
$$- \sinh 2 K_1 \cos \omega' - sinh2(K_2 + K_3 + K_4)$$
$$\times \left[\cos(\omega_x + \phi_x) + \cos(\omega_y + \phi_y)\right.$$
$$\left. + \cos(\omega_z + \phi_z)\right]\right] d\omega'd\omega_xd\omega_yd\omega_z$$

(30)

Here $\omega_x = \frac{2t_x\pi}{n}$, $\omega_y = \frac{2t_y\pi}{l}$ and $\omega_z = \frac{2t_z\pi}{o}$. In the formulates above (*i.e.*, Eqs. (29) and (30)), we have used the topological phases $\phi_x$, $\phi_y$ and $\phi_z$ to replace the weight factors $w_x$, $w_y$ and $w_z$ represented in Eq. (33) of [18].

The spontaneous magnetization $M$ of the 3D spinless fermionic model is the same as that obtained in [18] for the 3D Ising model:

$$M = \left[1 - \frac{16x_1^2 x_2^2 x_3^2 x_4^2}{(1-x_1^2)^2(1-x_2^2 x_3^2 x_4^2)^2}\right]^{\frac{3}{8}}$$

(31)

with $x_i = e^{-2K_i}$ ($i$ = 1, 2, 3, 4).

The true range $\kappa_x$ of the correlation of the 3D spinless fermionic model is determined by [18]:

$$[\kappa_x a]^{3/2} = 2(K_1^* - K_2 - K_3 - K_4)$$

(32)

Here $\kappa_x = 1/\xi$ with the correlation length $\xi$. It is known that the mass gap $m$ of the field theory corresponds to the reciprocal of the correlation length $\xi$ of the statistical mechanics, *i.e.*, $m = \frac{1}{\xi a}$ [13]. The mass gap $m$ is expressed by:

$$\Lambda_k = m = \left[\frac{2}{a^3}(K_1^* - K_2 - K_3 - K_4)\right]^{2/3}$$

(33)

To make a field theory with particles of small masses, the underlying statistical mechanics must be nearly critical. In particular, taking the continuum limit $a \to 0$ one reaches an interesting theory, with an excitation spectrum of finite masses, only if $\xi \to \infty$ (in the critical regions). Therefore, $K_1^* = K_2 + K_3 + K_4$ determines the critical point of the 3D spinless fermionic model, in consistent with the star-triangle relation for the duality between two dual lattices. When $K_1 = K_2 = K_3 = K$, the critical point is determined by the formula $K^* = 3K$, resulting in the golden ratio solution $x_c = e^{-2K_c} = \frac{\sqrt{5}-1}{2} = 0.6180339887...$, $K_c = 0.24060591...$ and $1/K_c = 4.15617384..$ [18]. The critical exponents of the 3D spinless fermionic model are equivalent to those for the 3D Ising model, which are in the universality class of $\alpha = 0$, $\beta = 3/8$, $\gamma = 5/4$, $\delta = 13/3$, $\eta = 1/8$ and $\nu = 2/3$ [18]. The experimental data [41,42] confirm the existence of the 3D Ising universality class in the 3D Ising magnets, which affirm the validity of the exact solutions of the 3D Ising models [18]. The Monte Carlo simulations on the 3D Ising structures illustrated in Figure 5 of ref. [21], which consist of two parts of contributions (local spin alignments and nonlocal long-range spin entanglements), gave the critical exponents fitting well with the exact solutions [43].

## 5. Dimensionality and Topological phases

5.1. Dimensionality

It is important to inspect the dimensionality of the present system. The original Hamiltonian $H_0$ in Eq. (1) is for the 3D Ising model. The interactions in the Hamiltonian $H_1$ in Eq. (2) are summed over lattice sites $N_1 = nl$ in a plane, which is reduced by applying the periodic condition. Keeping in mind the powers of eigenvalues $\lambda_{i,i'}^m$ for the partition function $Z$ [18,19] and the three interactions $K_1^*$, $K_2$ and $K_3$ in $H_1$, the system is still a 3D Ising model. Employing the Jordan-Wigner transformation does not change the dimensionality of the system, so the Hamiltonian $H_2$ in Eq. (5) is for the 3D model of spinless fermions. The model is equivalently viewed as a (2+1)D spinless fermionic model (time is the third dimension). Performing the time average within Jordan-von Neumann-Wigner framework and using the multiplication of Jordan algebra expand the system to be (3+1) dimensional. The sum of interactions in the Hamiltonian $H_3$ (Eq. (6)) takes over $N_2 = nlo$, where o represents the lattice points along the fourth direction. The linearization procedure and the local gauge transformation do not change the dimensionality of the system, so the Hamiltonians $H_4$ - $H_6$ in Eqs. (11), (14) and (18) are of (3+1) dimensions. During the Fourier transformation and Bogoliubov transformation, the dimensionality maintains so that the Hamiltonians $H_7$ and $H_8$ in Eqs. (20) and (25) are represented in the (3+1)D framework. The conservation of energy must be held by projecting the eigenvalues (Eq. (29)), the eigenvectors, the partition function (Eq. (30)) and the energy-momentum dispersion (Eq. (28)) of the (3+1)D system back to the 3D one by using the weights (*i.e.*, the topological phases) or by the dimension analysis. As mentioned in Section 2, the internal factors $W_j$ in the Hamiltonian $H_2$ (Eq. (5)) can be seen as the Chern-Simons gauge fields act on each site [16,37]. It has been proven that a 2D Ising model with a transverse field is equivalent to a 3D Ising model [22]. It is understandable that Chern-Simons gauge fields can be treated effectively as an additional dimension. It is worth noting that the critical exponents β = 3/8 and ν = 2/3 are determined by the dimension analyses on the spontaneous magnetization $M$ and the true range $κ_x$ of the correlation, upon the three vectors of the quaternionic eigenvectors of the 3D Ising model [18] and related models.

5.2. Topological phases

The local gauge transformation (*i.e.*, the topological Lorentz transformation) trivialize the nontrivial topological structures in the Hamiltonian on a trivial manifold into a Hamiltonian with the trivial topological structures on a nontrivial manifold for the eigenvectors. Meanwhile, three topological phases appear on the quaternionic eigenvectors of the 3D spinless fermionic model. The topological phases have a root for their origin, but uncovered in different features by different theories in mathematics and physics, which are connected to different effects in various fields. In this subsection, we shall discuss the phase factors in the 3D spinless fermionic model and their relationship with phases in other areas.

The phase factors are generated by a global effect of geometrical aspects of the system, which can be described by the Gauss-Bonet-Chern formula [44]:

$$\int_M Pf(\Omega)\, dM = (2\pi)^m \chi(M)$$

(34)

where $\Omega$ is the curvature of a $2m$-dimensional Riemann manifold $M$ given via the Levi-Civita connection, the Pfaffian of $\Omega$ is defined by $Pf(\Omega) = [\det(\Omega)]^{1/2}$ and $\chi(M)$ is the Chern number for the class $m$. The topological phases are determined by analyzing the geometric aspects with the Chern number of the first class ($m = 1$) $c_1 = 2$ [21]. The Röhrl Theorem [45] provides the possibility of the existence of a multi-valued function with regular singularities for a given monodromy representation, generating the topological phases. The global effect requires processing the gauge transformation between the spins at different sites along different dimensions. The gauge transformation for the 3D spinless fermionic model is similar to that in electromagnetism [46], Yang-Mills non-Abelian gauge theory [47] and Chern-Simons theory [37]. The concept of nonintegrable phase factors and global formulation of gauge fields have an identification with the mathematical concept of connections on principal fiber bundles [48]. The topological phases in the present spinless fermionic model have the origin analogous also to Aharonov-Bohm effect [49], Berry phase [50], Wilson loop [51,52] and quantum Hall effect [53-55], *etc*.

An effect of spins appears as a change in the Aharonov-Bohm period (in units of the flux quantum $\Phi_0 = hc/e$) [49]. In the present spinless fermionic system, an effect appears analogous to the Aharonov-Bohm effect, owing to the 3D many-body interacting spin lattice. The phase factor $e^{iA_{ij}(x)}$ generated by the magnetic flux $B(r)$ through an elementary plaquette *p* is given by [16]:

$$B(r) = \epsilon_{ij}\Delta_i A_j(x) \equiv \epsilon_{ij}\big(A_j(x + \hat{e}_i) - A_j(x)\big)$$

(35)

The effective magnetic field acts on the fermionic operators (the hopping term) as the phase factor $e^{iA_{ij}}$ of integrals of the vector potential *A* [31,34].

$$A_{ij} = 2\pi \int_i^j A \cdot dl$$

(36)

The phase factor can be written also in forms of the vector potential *A* and the interactions $J_{ij}$ [28,31]:

$$\exp(i2\pi\phi_p) = \prod AJ_{ij}$$

(37)

where the product is around the plaquette $p$, $\phi_p = (2\pi)^{-1} \sum_{\text{plaquette}} A_{ij}$ is the flux through each plaquette $p$ in units of flux quanta [34].

Next, we discuss the Berry phase described as [49]:

$$\phi = i\oint_C \langle \psi_n(r)|\nabla_r|\psi_n(r)\rangle \cdot dr = \oint_C A_{n,r} \cdot dr$$

(38)

with the Berry connection $A_{n,r}$. The concept of the Berry phase can be applied to the momentum space of energy bands with wave functions $\psi_n(k)$:

$$\phi = i\oint_C \langle \psi_n(k)|\nabla_k|\psi_n(k)\rangle \cdot dk = \oint_C A_{n,k} \cdot dk$$

(39)

with the Berry connection $A_{n,k}$. According to the Stokes theorem, the Berry phase in the momentum space is written as:

$$\phi = \int_{BZ} (\nabla_k \times A_{n,k}) \cdot dS_k$$

(40)

The Berry curvature $\nabla_k \times A_{n,k}$ corresponds to Pf($\Omega$) in the Gauss-Bonet-Chern formula (see Eq.(34)) [44] and we can determine the topological phases in the first Brillouin zone.

Then, we pay attention on the Wilson loop [51,52]. The chirality operator is defined as [32]:

$$W(C) = \text{Tr} \prod_{i \in C} \left(\frac{1}{2} + \sigma \cdot S_i\right)$$

(41)

where $S_i$ are Pauli matrices and $C$ is a lattice contour. The topological order parameter $W(C)$ acquires the form of a lattice Wilson loop [51,52].

$$W(C) = e^{i\phi(C)} = \prod_C e^{iA_{ij}}$$

(42)

which may be associated with the magnetic flux penetrating through a surface enclosed by the contour $C$, being represented by the gauge field $A_{ij}$ related with the phase/ spin fluctuation.

Finally, we are focused on the values for the topological phases. The topological phases were determined for the 3D Ising model by analysis of geometry [21], according to the Gauss-Bonnet-Chern formula [44]. $\phi_x = 2\pi$ is invariant for the property of the 2D Ising model, while $\phi_y$ and $\phi_z$ are responsible for the topological phases caused by the interaction along the third dimension [21]. The latter two phases together describe the long-rang entanglements between spins in a plane. An immediate consequence is that the topological phases of the 3D spinless fermion model are the same as those for the 3D Ising model, leading to $\phi_x = 2\pi$, $\phi_y = \pi/2$ and $\phi_z = \pi/2$. The same results can be obtained by analyzing the 3D spinless fermion model itself. As illustrated in Fig. 1 of ref. [33] for a 2D lattice, an arrow on the bond indicates a hopping along the direction carries a phase $e^{i\Phi/4} = e^{i\pi/4}$, since the flux through a plaquette is $\Phi = \pi$. This can be generated to a 3D lattice, so that the gauge transformation from the first site ($j + n$) of the third dimension to the first site ($j + 1$) of the second dimension, we need to account two arrows (along two bonds), thus the phase factor is $e^{i\Phi/2} = e^{i\pi/2}$, i.e., $\phi_y = \pi/2$. Furthermore, the gauge transformation from the first site ($j + n'$) of the fourth dimension (for the virtual lattice) to the first site ($j + 1$) of the second dimension has the same phase factor $e^{i\pi/2}$, i.e., $\phi_z = \pi/2$. The topological phases can be viewed in the following manner: In a manifold $S^1 \times S^1 \times S^1 \times R^1$ for the 3D Ising model [23] and also the 3D spinless fermionic model, the entanglement can be represented by a helix with a pitch $\boldsymbol{k'}$ with quaternionic components $k'_x$, $k'_y$ and $k'_z$, so that $\phi_x = 2k'_x r_x$, $\phi_y = 2k'_y r_y$ and $\phi_z = 2k'_z r_z$.

## 6. Conclusions

In conclusion, the exact solution of the 3D (or (2+1)D) spinless fermionic model is derived by performing a diagonalization process consisting of the Clifford algebraic approach, the Fourier transformation and the Bogoliubov transformation. The Clifford algebraic approach includes a time average within the Jordan-von Neumann-Wigner framework, a linearization procedure and a local gauge transformation (*i.e.*, a topological Lorentz transformation). The eigenvalues, the partition function, the subsequent thermodynamic properties and the critical behaviors are investigated for the 3D spinless fermionic model. The dimensionality and the topological phases are discussed. The advances in the 3D Ising model and related models not only provide a better understanding on the many-body interacting systems in condensed maters, but also benefit to solving the hard problems in mathematics and computer sciences

[56-61].

**Acknowledgements**

This work has been supported by the National Natural Science Foundation of China under grant number 52031014.

**References:**

[1] M. Ogata and H. Shiba, Bethe-ansatz wave function, momentum distribution, and spin correlation in the one-dimensional strongly correlated Hubbard model, Phys. Rev. B **41**, 2326-2338 (1990).

[2] N. Read and D. Green, Paired states of fermions in two dimensions with breaking of parity and time-reversal symmetries and the fractional quantum Hall effect, Phys. Rev. B **61**, 10267-10297 (2000).

[3] A.P. Schnyder, S. Ryu, A. Furusaki, and A.W.W. Ludwig, Classification of topological insulators and superconductors in three spatial dimensions, Phys. Rev. B **78**, 195125 (2008).

[4] L. Fu and C. L. Kane, Superconducting proximity effect and Majorana fermions at the surface of a topological insulator, Phys. Rev. Lett. **100**, 096407 (2008).

[5] S. Raghu, X.L. Qi, C. Honerkamp, and S.C. Zhang, Topological Mott insulators, Phys. Rev. Lett. **100**, 156401 (2008).

[6] K. Sun, H. Yao, E. Fradkin, and S.A. Kivelson, Topological insulators and Nematic phases from spontaneous symmetry breaking in 2D Fermi systems with a

quadratic band crossing, Phys. Rev. Lett. **103**, 046811 (2009).

[7] M. Neupane, I. Belopolski, M.M. Hosen, D.S. Sanchez, R. Sankar, M. Szlawska, S.Y. Xu, K. Dimitri, N. Dhakal, P. Maldonado, P.M. Oppeneer, D. Kaczorowski, F.C. Chou, M.Z. Hasan, and T. Durakiewicz, Observation of topological nodal fermion semimetal phase in ZrSiS, Phys. Rev. B **93**, 201104(R) (2016).

[8] E. Ising, Beitrag zur theorie des ferromagnetismus, Z Phys, **31**, 253-258 (1925).

[9] T.D. Schultz, D.C. Mattis, and E.H. Lieb, Two-dimensional Ising model as a soluble problem of many fermions. Rev. Mod. Phys., **36**, 856-871 (1964).

[10] L. Onsager, Crystal Statistics I: A two-dimensional model with an order-disorder transition, Phys. Rev. **65**, 117-149 (1944).

[11] B. Kaufman, Crystal Statistics II: Partition function evaluated by spinor analysis, Phys. Rev. **76**, 1232-1243 (1949).

[12] P. Jordan and E. Wigner, Über das Paulische Äquivalenzverbot, Z. Phys. **47**, 631-651 (1928).

[13] J.B. Kogut, An introduction to lattice gauge theory and spin systems, Rev. Mod. Phys. **51**, 659-713 (1979).

[14] F.D.M. Haldane, General relation of correlation exponents and spectral properties of one-dimensional Fermi systems: Application to the anisotropic S = 1/2 Heisenberg chain, Phys. Rev. Lett. **45**, 1358-1362 (1980).

[15] G. Bouzerar, D. Poilblanc, and G. Montambaux, Persistent currents in one-dimensional disordered rings of interacting electrons, Phys. Rev. B **49**, 8258-8262 (1994).

[16] E. Fradkin, Jordan-Wigner transformation for quantum-spin systems in two dimensions and fractional statistics, Phys. Rev. Lett. **63**, 322-325 (1989).

[17] E. Fradkin, M. Srednicki and L. Susskind, Fermion representation for the $Z_2$ lattice gauge theory in 2+1 dimensions, Phys. Rev. D **21**, 2885-2891 (1980).

[18] Z.D. Zhang, Conjectures on the exact solution of three - dimensional (3D) simple orthorhombic Ising lattices, Phil. Mag. **87**, 5309-5419 (2007).

[19] Z.D. Zhang, O. Suzuki, and N.H. March, Clifford algebra approach of 3D Ising model, Advances in Applied Clifford Algebras, **29**, 12 (2019).

[20] O. Suzuki and Z.D. Zhang, A method of Riemann-Hilbert problem for Zhang's conjecture 1 in a ferromagnetic 3D Ising model: trivialization of topological structure, Mathematics, **9**, 776 (2021).

[21] Z.D. Zhang and O. Suzuki, A method of the Riemann-Hilbert problem for Zhang's conjecture 2 in a ferromagnetic 3D Ising model: topological phases, Mathematics, **9**, 2936 (2021).

[22] Z.D. Zhang, Exact solution of two-dimensional (2D) Ising model with a transverse field: a low-dimensional quantum spin system, Physica E **128**, 114632 (2021).

[23] Z.D. Zhang, Exact solution of the three-dimensional (3D) $Z_2$ lattice gauge theory, Open Physics **23**, 20250215 (2025).

[24] Z.D. Zhang, Mathematical basis, phase transitions and singularities of (3+1)-dimensional $\phi^4$ scalar field model, arXiv: 2511.07439.

[25] P. Jordan, J. von Neumann, and E. Wigner, On an algebraic generalization of the

quantum mechanical formalism. Ann. of Math. **35**, 29-64 (1934).

[26] W.F. Li, L.T. Ding, R. Yu, T. Roscilde, and S. Haas, Scaling behavior of entanglement in two- and three-dimensional free-fermion systems, Phys. Rev. B **74**, 073103 (2006).

[27] Y.R. Wang, Ground state of the two-dimensional antiferromagnetic Heisenberg model studied using an extended Wigner-Jordon transformation, Phys. Rev. B **43**, 3786-3789(1991).

[28] J. Klinovaja and D. Loss, Fractional fermions with non-Abelian statistics, Phys. Rev. Lett. **110**, 126402 (2013).

[29] B. Bock and M. Azzouz, Generalization of the Jordan-Wigner transformation in three dimensions and its application to the Heisenberg bilayer antiferromagnet, Phys. Rev. B **64**, 054410 (2001).

[30] B. Olmos, R. González-Férez, and I. Lesanovsky, Creating collective many-body states with highly excited atoms, Phys. Rev. A **81**, 023604 (2010).

[31] W. Barford and J. H. Kim, Spinless fermions on frustrated lattices in a magnetic field, Phys. Rev. B **43**, 559-568 (1991).

[32] B. Basu, S. Dhar, and P. Bandyopadhyay, High-$T_c$ superconductivity, skyrmions, and the Berry phase, Phys. Rev. B **69**, 094505 (2004).

[33] J.P. Lu and W. Barford, Pairing of spinless fermions in two dimensions, Phys. Rev. B **44**, 5263-5268 (1991).

[34] Y. Tan and D. J. Thoules, Total energy for fermions on a two-dimensional lattice in a magnetic field, Phys. Rev. B **46**, 2985-2994 (1992).

[35] M. Suzuki, Relationship among exactly soluble models of critical phenomena, -- 2D Ising model, dimer problem and the generalized XY-model, Progress Theor. Phys. **46**, 1337-1359 (1971).

[36] A. De Pasquale and P. Facchi, *XY* model on the circle: Diagonalization, spectrum, and forerunners of the quantum phase transition, Phys. Rev. A **80**, 032102 (2009).

[37] S.S. Chern and J. Simons, Characteristic forms and geometric invariants, Annals Math. **99**, 48-69 (1974).

[38] Z.D. Zhang, Topological quantum statistical mechanics and topological quantum field theories, Symmetry **14**, 323 (2022).

[39] S.L. Lou and S.H. Wu, Three-dimensional Ising model and transfer matrices. Chin. J. Phys. **38**, 841-854 (2000)

[40] Z.D. Zhang, Mathematical structure of the three - dimensional (3D) Ising model, Chinese Phys. B **22**, 030513 (2013).

[41] Z.D. Zhang and N.H. March, Three - dimensional (3D) Ising universality in magnets and critical indices at fluid-fluid phase transition, Phase Transitions **84**, 299-307 (2011).

[42] Z.D. Zhang, Universality of critical behaviors in the three-dimensional (3D) Ising magnets, arXiv: 2510.09111.

[43] B.C. Li and W. Wang, Exploration of dynamic phase transition of 3D Ising model with a new long-range interaction by using the Monte Carlo method, Chinese Journal of Physics **90**, 15-30 (2024).

[44] S.S. Chern, On the curvatura integra in a Riemannian manifold. Ann. Math., **46**,

674-684 (1945).

[45] H. Röhrl, Das Riemannsch-Hilbertsche problem der theorie der linieren differentialgleichungen, Math. Ann. **133**, 1-25 (1957).

[46] T.W. Barrett, Topological Foundations of Electromagnetism, (World Scientific, Singapore, 2008).

[47] C.N. Yang and R.L. Mills, Conservation of isotopic spin and isotopic gauge invariance, Phys. Rev. **96**, 191-195 (1954).

[48] T.T. Wu and C.N. Yang, Concept of nonintegrable phase factors and global formulation of gauge fields. Phys. Rev. **D 12**, 3845-3857 (1975).

[49] Y. Aharonov and D. Bohm, Significance of electromagnetic potentials in the quantum theory, Phys. Rev. **115**, 485-491 (1959).

[50] M.V. Berry, Quantal phase-factors accompanying adiabatic changes, Proc. R, Soc. London, A **392**, 45-57 (1984).

[51] K.G. Wilson, Confinement of qnarks, Phys. Rev. D **10**, 2445-2459 (1974).

[52] Y.M. Makeenko, Brief introduction to Wilson loops and large N, Physics of Atomic Nuclei, **73**, 878-994 (2010).

[53] F.D.M. Haldane, Model for a quantum Hall effect without Landau levels: Condensed-matter realization of the "parity anomaly", Phys. Rev. Lett. **61**, 2015-2018 (1988).

[54] Q. Niu, D.J. Thouless, and Y.S. Wu, Quantized Hall conductance as a topological invariant, Phys. Rev. B**31**, 3372-3377 (1985).

[55] C.L. Kane and E.J. Mele, Quantum spin Hall effect in graphene, Phys. Rev. Lett.

**95**, 226801 (2005).

[56] Z.D. Zhang, Computational complexity of spin-glass three-dimensional (3D) Ising model, J. Mater. Sci. Tech. **44**, 116-120 (2020).

[57] Z.D. Zhang, Mapping between spin-glass three-dimensional (3D) Ising model and Boolean satisfiability problems, Mathematics **11**, 237 (2023).

[58] Z.D. Zhang, Lower bound of computational complexity of knapsack problems, AIMS Math. **10**, 11918-11938 (2025).

[59] Z.D. Zhang, Relationship between spin-glass three-dimensional (3D) Ising model and traveling salesman problems, arXiv: 2507.01914.

[60] Z.D. Zhang, Equivalence between the zero distributions of the Riemann zeta function and a two-dimensional Ising model with randomly distributed competing interactions, arXiv: 2411.16777.

[61] Z.D. Zhang, Equivalence between the pair correlation functions of primes and of spins in a two-dimensional Ising model with randomly distributed competing interactions, arXiv: 2508.14938.